%% file: revised.tex
\documentclass[conference]{IEEEtran}
\IEEEoverridecommandlockouts
\usepackage{cite}
\usepackage{amsmath,amssymb,amsfonts}
\usepackage{algorithmic}
\usepackage{graphicx}
\usepackage{textcomp}
\usepackage{xcolor}
\def\BibTeX{{\rm B\kern-.05em{\sc i\kern-.025em b}\kern-.08em
		T\kern-.1667em\lower.7ex\hbox{E}\kern-.125emX}}

\begin{document}
	
\title{Power Allocation in Multi-user Cellular Networks With Deep Q Learning Approach}

\author{Fan~Meng, Peng~Chen and Lenan~Wu
	\thanks{Fan~Meng, and Lenan~Wu are with the School of Information Science and
		Engineering, Southeast University, Nanjing 210096, China (e-mail: mengxiaomaomao@outlook.com, wuln@seu.edu.cn).}
	\thanks{
		Peng Chen is with the State Key Laboratory of Millimeter Waves, Southeast University, Nanjing 210096, China (e-mail: chenpengseu@seu.edu.cn).}
}

\maketitle

\begin{abstract}
	
The model-driven power allocation (PA) algorithms in the wireless cellular networks with interfering multiple-access channel (IMAC) have been investigated for decades. Nowadays, the data-driven model-free machine learning-based approaches are rapidly developed in this field, and among them the deep reinforcement learning (DRL) is proved to be of great promising potential. Different from supervised learning, the DRL takes advantages of exploration and exploitation to maximize the objective function under certain constraints. In our paper, we propose a two-step training framework. First, with the off-line learning in simulated environment, a deep Q network (DQN) is trained with deep Q learning (DQL) algorithm, which is well-designed to be in consistent with this PA issue. Second, the DQN will be further fine-tuned with real data in on-line training procedure. The simulation results show that the proposed DQN achieves the highest averaged sum-rate, comparing to the ones with present DQL training. With different user densities, our DQN outperforms benchmark algorithms and thus a good generalization ability is verified.

\end{abstract}

\begin{IEEEkeywords}
Deep reinforcement learning, deep Q learning, interfering multiple-access channel, power allocation.
	
\end{IEEEkeywords}

\section{Introduction}\label{sec:intro}

Data transmitting in wireless communication networks has experienced explosively growth in recent decades and will keep rising in the future. The user density is greatly increasing, resulting in critical demand for more capacity and spectral efficiency. Therefore, both intra-cell and inter-cell interference managements are significant to improve the overall capacity of a cellular network system. The problem of maximizing a generic sum-rate is studied in this paper, and it is non-convex, NP-hard and cannot be solved efficiently.

Various model-driven algorithms have been proposed in the present papers for PA problems, such as fractional programming (FP)~\cite{Shen2018Fractional}, weighted MMSE (WMMSE)~\cite{Shi2011An} and some others~\cite{Chiang2008Power, 5770666}. Excellent performance can be observed through theoretical analysis and numerical simulations, but serious obstacles are faced in practical deployments~\cite{Liye}. First, these techniques highly rely on tractable mathematical models, which are imperfect in real communication scenarios with the specific user distribution, geographical environment, etc. Second, the computational complexities of these algorithms are high.

In recent years, the machine learning (ML)-based approaches have been rapidly developed in wireless communications~\cite{8054694}. These algorithms are usually model-free, and are compliant with optimizations in practical communication scenarios. Additionally, with developments of graphic processing unit (GPU) or specialized chips, the executions can be both fast and energy-efficient, which brings in solid foundations for massive applications.

Two main branches of ML, supervised learning and reinforcement learning (RL)~\cite{Lecun2015Deep}, are briefly introduced here. With supervised learning, a deep neural network (DNN) is trained to approximate some given optimal (or suboptimal) objective algorithms, and it has been realized in some applications~\cite{8444648, 8454504, 8052521}. However, the target algorithm is usually unavailable and the performance of DNN is bounded by the supervisor. Therefore, the RL has received widespread attention, due to its nature of interacting with an unknown environment by exploration and exploitation. The Q learning method is the most well-studied RL algorithm, and it is exploited to cope with power allocation (PA) in~\cite{DBLP:journals/corr/abs-1803-06760, 7997440, 8466370}, and some others~\cite{8412128}. The DNN trained with Q learning is called deep Q network (DQN), and it is proposed to address the distributed downlink single-user PA problem~\cite{DBLP:DRL}. 

In our paper, we extend the work in~\cite{DBLP:DRL}, and the PA problem in cellular cells with multiple users is investigated. The design of the DQN model is discussed and introduced. Simulation results show that our DQN outperforms the present DQNs and the benchmark algorithms. The contributions of this work are summarized as follows:
\begin{itemize}
	\item A model-free two-step training framework is proposed. The DQN is first off-line trained with DRL algorithm in simulated scenarios. Second, the learned DQN can be further dynamically optimized in real communication scenarios, with the aid of transfer learning.
	\item The PA problem using deep Q learning (DQL) is discussed, then a DQN enabled approach is proposed to be trained with current sum-rate as reward function, including no future reward. The input features are well-designed to help the DQN get closer to the optimal solution. 
	\item After centralized training, the proposed DQN is tested by distributed execution. The averaged rate-sum of DQN outperforms the model-driven algorithms, and also shows good generalization ability in a series of benchmark simulation tests. 	
\end{itemize}

The remainder of this paper is organized as follows. Section~\ref{sec:system} outlines the PA problem in the wireless cellular network with IMAC. In Section~\ref{sec:dqn} our proposed DQN is introduced in detail. Then, this DQN is tested in distinct scenarios, along with benchmark algorithms, and the simulation results are analyzed in Section~\ref{sec:sim}. Conclusions and discussion are given in Section~\ref{sec:con}. 

\section{System Model}\label{sec:system}

\begin{figure}
	\centering
	\includegraphics[width=3.6in]{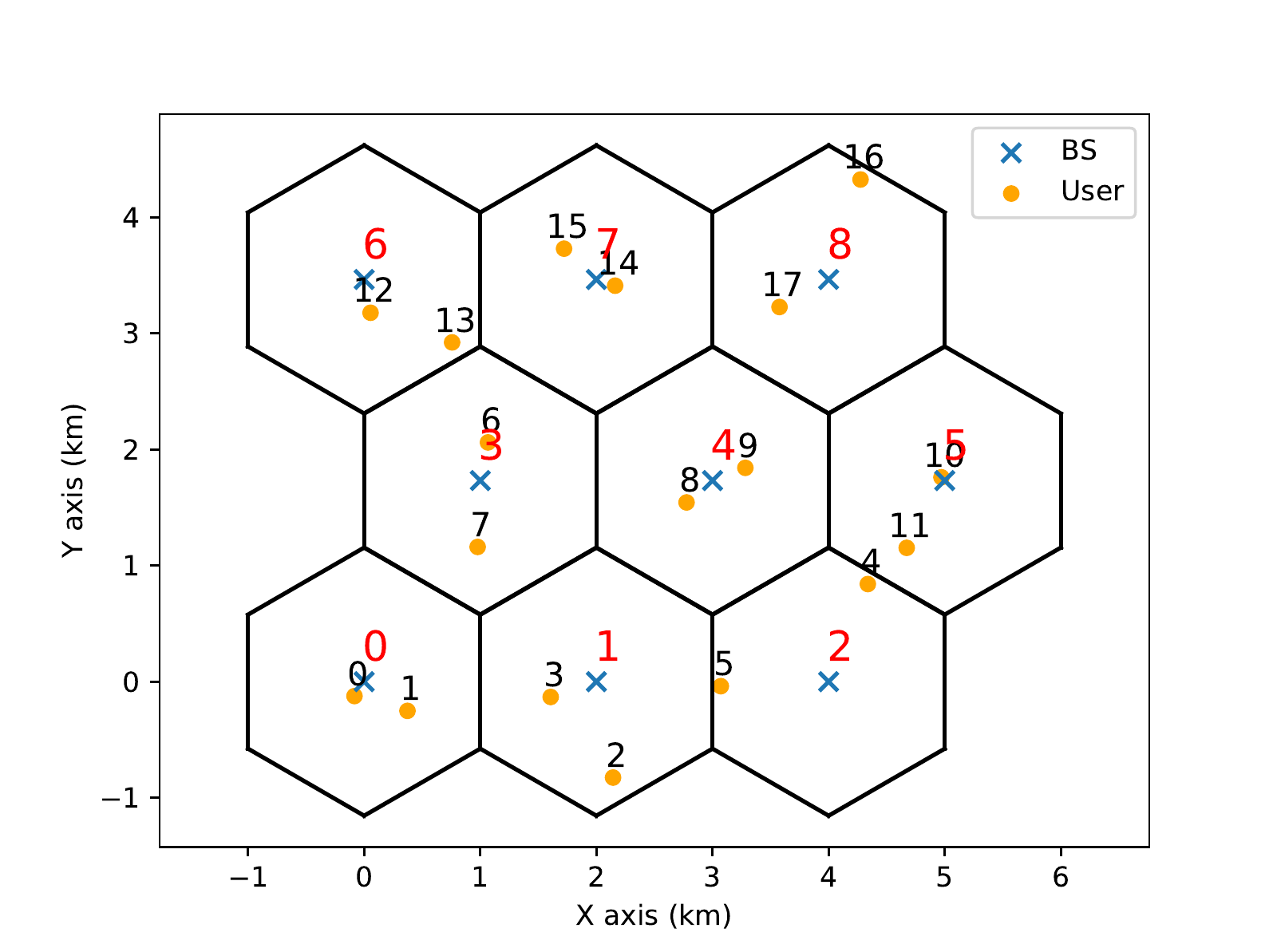}
	\caption{An illustrative example of a multi-user cellular network with $ 9 $ cells. In each cell, a BS serves $ 2 $ users simultaneously.}
	\label{fig:cellular}
\end{figure}

The problem of PA in the cellular network with interfering multiple-access channel (IMAC) is considered. In a system with $ N $ cells, at the center of each cell a base stations (BS) simultaneously serves $ K $ users with sharing frequency bands. A simple network example is shown in Fig.~\ref{fig:cellular}. At time slot $ t $, the independent channel coefficient between the $ n $-th BS and the user $ k $ in cell $ j $ is denoted by $ g^t_{n,j,k} $, and can be expressed as 
\begin{equation}\label{equ:g}
g_{n,j,k} = |h^t_{n,j,k}|^2\beta_{n,j,k},
\end{equation}
where $ h^t_{n,j,k} $ is the small scale complex flat fading element, and $ \beta_{n,j,k} $ is the large scale fading component taking account of both the geometric attenuation and the shadow fading. Therefore, the signal to interference plus noise ratio (SINR) of this link can be described by
\begin{equation}\label{equ:sinr}
\textup{sinr}^t_{n,k} = \frac{g^t_{n,n,k} p^t_{n,k}}{\sum_{k' \neq k} g^t_{n,n,k} p^t_{n,k'} + \sum_{n' \in D_n} g^t_{n',n,k} \sum_{j} p^t_{n',j} + \sigma^2},
\end{equation}
where $ D_n $ is the set of interference cells around the $ n $-th cell, $ p $ is the emitting power of BS, and $ \sigma^2 $ denotes the additional noise power. With normalized bandwidth, the downlink rate of this link is given as 
\begin{equation}\label{equ:C}
C^t_{n,k} = \log_2\left(1 + \textup{sinr}^t_{n,k}\right),
\end{equation}
The optimization target is to maximize this generic sum-rate objective function under maximum power constraint, and it is formulated as
\begin{equation}\label{equ:tar}
\begin{split}
&\max_{\textbf{p}^t}\quad \sum_n\sum_k C^t_{n,k}\\
&\textup{s.t.}\quad 0 \leq p^t_{n,k} \leq P_{\textup{max}}, \;\forall n,k,
\end{split}
\end{equation}
where $ \textbf{p}^t = \{p^t_{n,k}|\forall n,k\} $, and $ P_{\textup{max}} $ denotes the maximum emitting power. We also define sum-rate $ C^t = \sum_n\sum_k C^t_{n,k} $, $ \textbf{C}^t = \{C^t_{n,k}|\forall n,k\} $, and channel state information (CSI) $ \textbf{g}^t = \{g^t_{n,j,k}|\forall n,j,k\} $. This problem is non-convex and NP-hard, so we propose a data-driven learning algorithm based on the DQN model in the following section. 

\section{Deep Q Network}\label{sec:dqn}

\subsection{Background}
Q learning is one of the most popular RL algorithms aiming to deal with the Markov decision process (MDP) problems~\cite{Mnih2015Human}. At time instant $ t $, by observing the state $ s^t \in S $, the agent takes action $ a^t \in A $ and interacts with the environment, and then get the reward $ r^t $ and the next state $ s^{t+1} $ is obtained. The notations $ A $ and $ S $ are the action set and the state set, respectively. Since $ S $ can be continuous, the DQN is proposed to combine Q learning with a flexible DNN to settle infinite state space. The cumulative discounted reward function is given as
\begin{equation}\label{equ:Rt}
R^t = \sum_{\tau = 0}^{\infty} \gamma^{\tau} r^{t+\tau+1},
\end{equation}
where $ \gamma \in [0, 1) $ is a discount factor that trades off the importance of immediate and future rewards, and $ r $ denotes the reward. Under a certain policy $ \pi $, the Q function of the agent with an action $ a $ in state $ s $ is given as
\begin{equation}\label{equ:value}
Q_{\pi}(s,a;\boldsymbol{\theta}) = \mathbb{E}_{\pi}\left[R^t | s^t = s, a^t = a\right],
\end{equation}
where $ \boldsymbol{\theta} $ denotes the DQN parameters, and $ \mathbb{E}\left[\cdot\right] $ is the expectation operator. Q learning concerns with how agents ought to interact with an unknown environment so as to maximize the Q function. The maximization of~\eqref{equ:value} is equivalent to the Bellman optimality equation~\cite{Introduction}, and it is describe as
\begin{equation}\label{equ:yt}
y^t = r^{t} + \gamma \max_{a'} Q(s^{t+1},a';\boldsymbol{\theta}^t),
\end{equation}
where $ y^t $ is the optimal Q value. The DQN is trained to approximate the Q function, and the standard Q learning update of the parameters $ \boldsymbol{\theta} $ is described as
\begin{equation}\label{equ:update}
\boldsymbol{\theta}^{t+1} = \boldsymbol{\theta}^{t} + \eta \left(y^t - Q(s^t,a^t;\boldsymbol{\theta}^t)\right)\nabla Q(s^t,a^t;\boldsymbol{\theta}^t),
\end{equation}
where $ \eta $ is the learning rate. This update resembles stochastic gradient descent, gradually updating the current value $ Q(s^{t},a^{t};\boldsymbol{\theta}^t) $ towards the target $ y^t $. The experience data of the agent is loaded as $ \left(s^t, a^t, r^t, s^{t+1}\right) $. The DQN is trained with recorded batch data randomly sampled from the experience replay memory, which is a first-in first-out queue.

\subsection{Discussion on DRL}\label{sec:dis}

In many applications such as playing video games~\cite{Mnih2015Human}, where current strategy has long-term impact on cumulative reward, the DQN achieve remarkable results and beat humans. However, the discount factor is suggested to be zero in this PA problem. The DQL aims to maximize the Q function. Let $ \gamma = 0 $, from~\eqref{equ:value} we have
\begin{equation}\label{equ:q1}
\max Q = \max_{a \in A} \mathbb{E}_{\pi}\left[r^t | s^t = s, a^t = a\right].
\end{equation}
For a PA problem, clearly that $ s = \textbf{g}^t $, $ a = \textbf{p}^t $. Then we let $ r^t = C^t $ and get that
\begin{equation}\label{equ:q2}
\max Q = \max_{\textbf{0} \preceq \textbf{p}^t \preceq \textbf{p}_{\max}} \mathbb{E}_{\pi}\left[C^t | \textbf{g}^t, \textbf{p}^t\right].
\end{equation}
In the execution period the policy is deterministic, and thus~\eqref{equ:q2} can be written as 
\begin{equation}\label{equ:q3}
\max Q = \max_{\textbf{0} \preceq \textbf{p}^t \preceq \textbf{p}_{\max}} C^t \left(\textbf{g}^t, \textbf{p}^t\right),
\end{equation}
which is a equivalent form of~\eqref{equ:tar}. In this inference process we assume that $ \gamma = 0 $ and $ r^t = C^t $, indicating that the optimal solution to~\eqref{equ:tar} is identical to that of~\eqref{equ:value}, under these two conditions.

\begin{figure}
	\centering
	\includegraphics[width=3.6in]{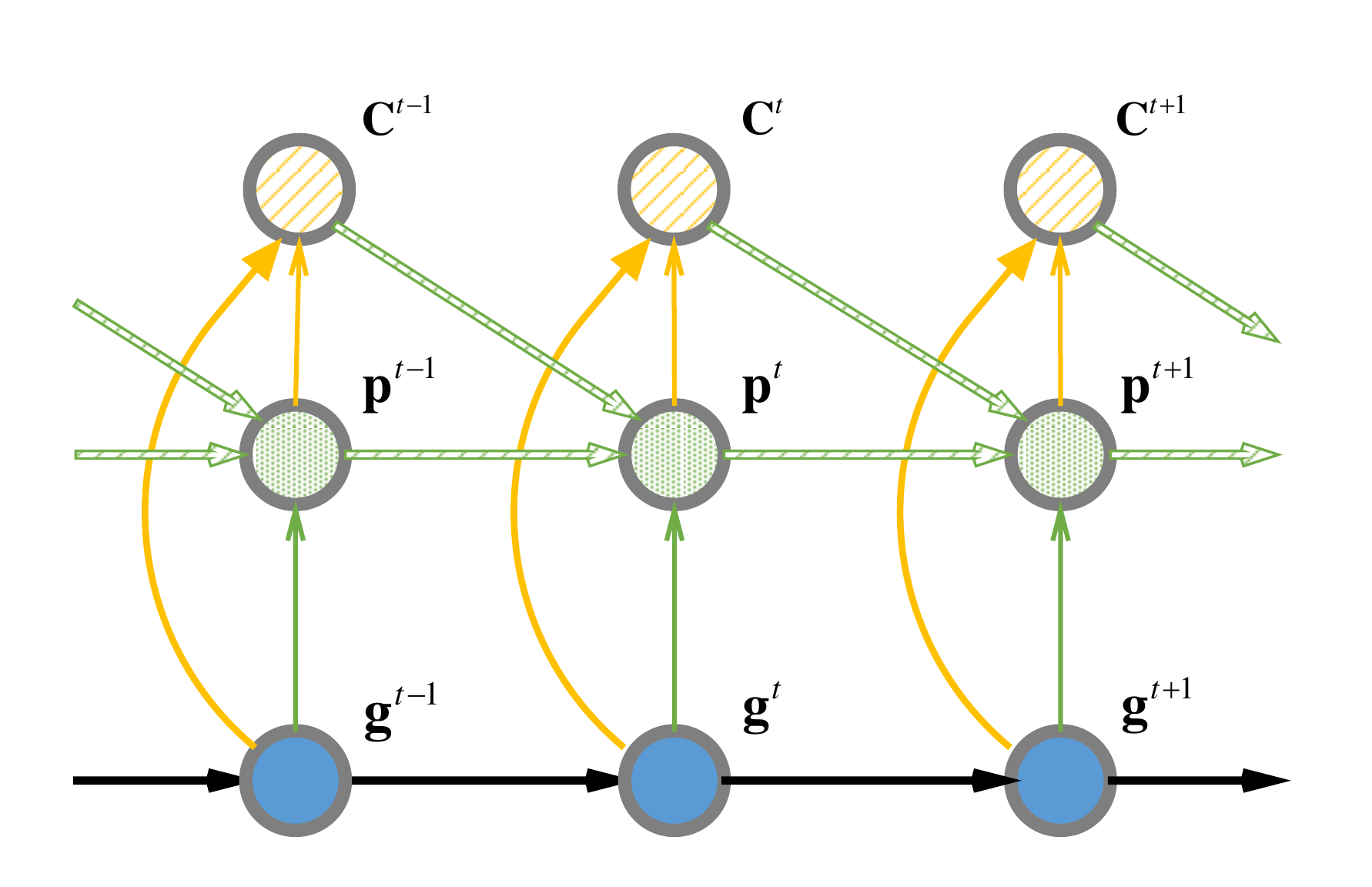}
	\caption{The solution of DQN is determined by CSI $ \textbf{g}^t $, along with downlink rate $ \textbf{C}^{t-1} $ and transmitting power $ \textbf{p}^{t-1} $.}
	\label{fig:opt}
\end{figure}

As shown in Fig.~\ref{fig:opt}, it is well-known that the optimal solution $ \textbf{p}^{t*} $ of~\eqref{equ:tar} is only determined by current CSI $ \textbf{g}^t $, and the sum-rate $ \textbf{C}^t $ is calculated with $ (\textbf{g}^t, \textbf{p}^t) $. Theoretically the optimal power $ \textbf{p}^{t*} $ can be obtained using a DQN with input being just $ \textbf{g}^t $. In fact, the performance of this designed DQN is poor, since it is non-convex and the optimal point is hard to find. Therefore, we propose to utilize two more auxiliary features: $ \textbf{C}^{t-1} $ and $ \textbf{p}^{t-1} $. Since that the channel can be modeled as a first-order Markov process, the solution of last time period can help the DQN get closer to the optimum, and~\eqref{equ:q3} can be rewritten as
\begin{equation}\label{equ:q4}
\max Q = \max_{\textbf{0} \preceq \textbf{p}^t \preceq \textbf{p}_{\max}} C^t\left(\textbf{g}^t, \textbf{p}^t, \textbf{C}^{t-1}, \textbf{p}^{t-1}\right).
\end{equation}

Once $ \gamma = 0 $ and $ r^t = C^t $,~\eqref{equ:yt} is simplified to be $ y^t = C^t $, and the replay memory is also reduced to be $ \left(s^t, a^t, r^t\right) $. The DQN works as an estimator to predict the current sum-rate of corresponding power levels with a certain CSI. These discussions provide good guidance for the following DQN design.

\subsection{DQN Design in Cellular Network}

In our proposed model-free two-step training framework, the DQN is first off-line pre-trained with DRL algorithm in simulated wireless communication system. This procedure is to reduce the on-line training stress, due to the large data requirement of data-driven algorithm by nature. Second, with the aid of transfer learning, the learned DQN can be further dynamically fine-tuned in real scenarios. Since the practical wireless communication system is dynamic and influenced by unknown issues, the data-driven algorithm is believed to be a promising technique. We just discuss the two-step framework here, and the first training step is mainly focused in the following manuscript.

In a certain cellular network, each BS-user link is regarded as an agent and thus a multi-agent system is studied. However, multi-agent training is difficult since it needs much more learning data, training time and DNN parameters. Therefore, centralized training is considered, and only one agent is trained by using all agents' experience replay memory. Then, this agent's learned policy is shared in the distributed execution period. For our designed DQN, components of the replay memory are introduced as follows.

\subsubsection{State}

The state design for a certain agent $ (n,k) $ is important, since the full environment information is redundant and irrelevant elements must be removed. The agent is assumed to have corresponding perfect instant CSI information in~\eqref{equ:sinr}, and we define logarithmic normalized interferer set $ \boldsymbol{\Gamma}_{n,k}^t $ as
\begin{equation}\label{equ:gamma_g}
\boldsymbol{\Gamma}_{n,k}^t = \left\{ \underbrace{{1,\cdots,1}}_{K-1}, \left\{\log_2\left(1 + \frac{g^t_{n',j,k}}{g^t_{n,k,k}} \right)\bigg| n'\in D_n, \forall j \right\}\right\}.
\end{equation}
The channel amplitude of interferers are normalized by that of the needed link, and the logarithmic representation is preferred since the amplitudes of channel often vary by orders of magnitude. The cardinality of $ \boldsymbol{\Gamma}_{n,k}^t $ is $ (|D_n|+1)K-1 $. To further decrease the input dimension and reduce the computational complexities, the elements in $ \boldsymbol{\Gamma}_{n,k}^t $ are sorted in decrease turn and only the first $ C $ elements remain. As we discussed in~\ref{sec:dis}, these remained components' and this link's corresponding downlink rate $ \textbf{C}_{n,k}^{t-1} $ and transmitting power $ \textbf{p}_{n,k}^{t-1} $ at last time slot, are the additional two parts of the input to our DQN. Therefore, the state is composed of three features: $ s^t_{n,k} = \{ \boldsymbol{\Gamma}_{n,k}^t, \textbf{C}_{n,k}^{t-1}, \textbf{p}_{n,k}^{t-1} \} $. The cardinality of state, i.e., the input dimension for DQN is $ |S| = 3C+2 $. 

\subsubsection{Action}

In~\eqref{equ:tar} the downlink power is a continuous variable, and is only constrained by maximum power constraint. Since the action space of DQN must be finite, the possible emitting power is quantized in $ |A| $ levels. The allowed power set is given as
\begin{equation}\label{equ:p_set}
A = \left \{0, P_{\textup{min}}, P_{\textup{min}}\left(\frac{P_{\textup{max}}}{P_{\textup{min}}}\right)^{\frac{1}{|A|-2}}, \cdots, P_{\textup{max}}  \right\},
\end{equation}
where $ P_{\textup{min}} $ is the non-zero minimum emitting power.

\subsubsection{Reward}

In some manuscripts the reward function is elaborately designed to improve the agent's transmitting rate and also mitigate the interference influence. However, most of these reward functions are suboptimal approaches to the target function of~\eqref{equ:tar}. In our paper, the $ C^t $ is directly used as the reward function, and it is shared by all agents. In the training simulations with small or medium scale cellular network, this simple method proves to be feasible.


\section{Simulation Results}\label{sec:sim}

\subsection{Simulation Configuration}

A cellular network with $ N = 25 $ cells is simulated. At center of each cell, a BS is deployed to synchronously serve $ K = 4 $ users which are located uniformly and randomly within the cell range $ r \in [R_{\min}, R_{\min}] $, where $ R_{\min} = 0.01 $ km and $ R_{\min} = 1 $ km are the inner space and half cell-to-cell distance, respectively. The small-scale fading is simulated to be Rayleigh distributed, and the Jakes model is adopted with Doppler frequency $ f_d = 10 $ Hz and time period $ T = 20 $ ms. According to the LTE standard, the large-scale fading is modeled as $ \beta = 120.9 + 37.6 \log_{10}(d) + 10 \log_{10}(z) $ dB, where $ z $ is a log-normal random variable with standard deviation being $ 8 $ dB, and $ d $ is the transmitter-to-receiver distance (km). The AWGN power $ \sigma^2 $ is $ -114 $ dBm, and the emitting power constraints $ P_{\textup{min}} $ and $ P_{\textup{max}} $ are $ 5 $ and $ 38 $ dBm, respectively. 

A four-layer feed-forward neural network (FNN) is chosen as DQN, and the neuron numbers of two hidden layers are $ 128 $ and $ 64 $, respectively. The activation function of output layer is linear, and the ReLU is adopted in the hidden layers. The cardinality of adjacent cells is $ |D_n| = 18, \forall n $, the first $ C = 16 $ interferers remain and power level number $ |A| = 10 $. Therefore, the input and output dimensions are $ 50 $ and $ 10 $, respectively. 

In the off-line training period, the DQN is first randomly initialized and then trained epoch by epoch. In the first $ 100 $ episodes, the agents only take actions stochastically, then they follow by adaptive $ \epsilon $-greedy learning strategy~\cite{Introduction} to step in the following exploring period. In each episode, the large-scale fading is invariant, and thus the number of training episode must be large enough to overcome the generalization problem. There are $ 50 $ time slots per episode, and the DQN is trained with $ 256 $ random samples in the experience replay memory every $ 10 $ time slots. The Adam algorithm~\cite{Kingma2014Adam} is adopted as the optimizer in our paper, and the learning rate $ \eta $ exponentially decays from $ 10^{-3} $ to $ 10^{-4} $. All training hyper-parameters are listed in Tab.\ref{table:hyperparameter} for better illustration. In the following simulations, these default hyper-parameters will be clarified once changed.

The FP algorithm, WMMSE algorithm, maximum PA and random PA schemes are treated as benchmarks to evaluate our proposed DQN-based algorithm. The perfect CSI of current moment is assumed to be known for all schemes. The simulation code will be available after formal publication.

\begin{table}
	\caption{Hyper-parameters setup of DQN training}
	\centering
	\begin{tabular}{c|c||c|c}
		\hline
		\textbf{Parameter} & \textbf{Value} & \textbf{Parameter} & \textbf{Value}\\
		\hline
		Number of $ T $ per episode & $ 50 $ & Initial $ \eta $ & $ 10^{-3} $\\
		Observe episode number & $ 100 $ & Final $ \eta $ & $ 10^{-4} $\\
		Explore episode number & $ 9900 $ & Initial $ \epsilon $ & $ 0.2 $\\
		Train interval & 10  & Final $ \epsilon $ & $ 10^{-4} $\\
		Memory size & $ 50000 $ & Batch size & $ 256 $\\	 
		\hline
	\end{tabular}
	\label{table:hyperparameter}
\end{table}

\subsection{Discount Factor}\label{gamma}

\begin{figure}
	\centering
	\includegraphics[width=3.6in]{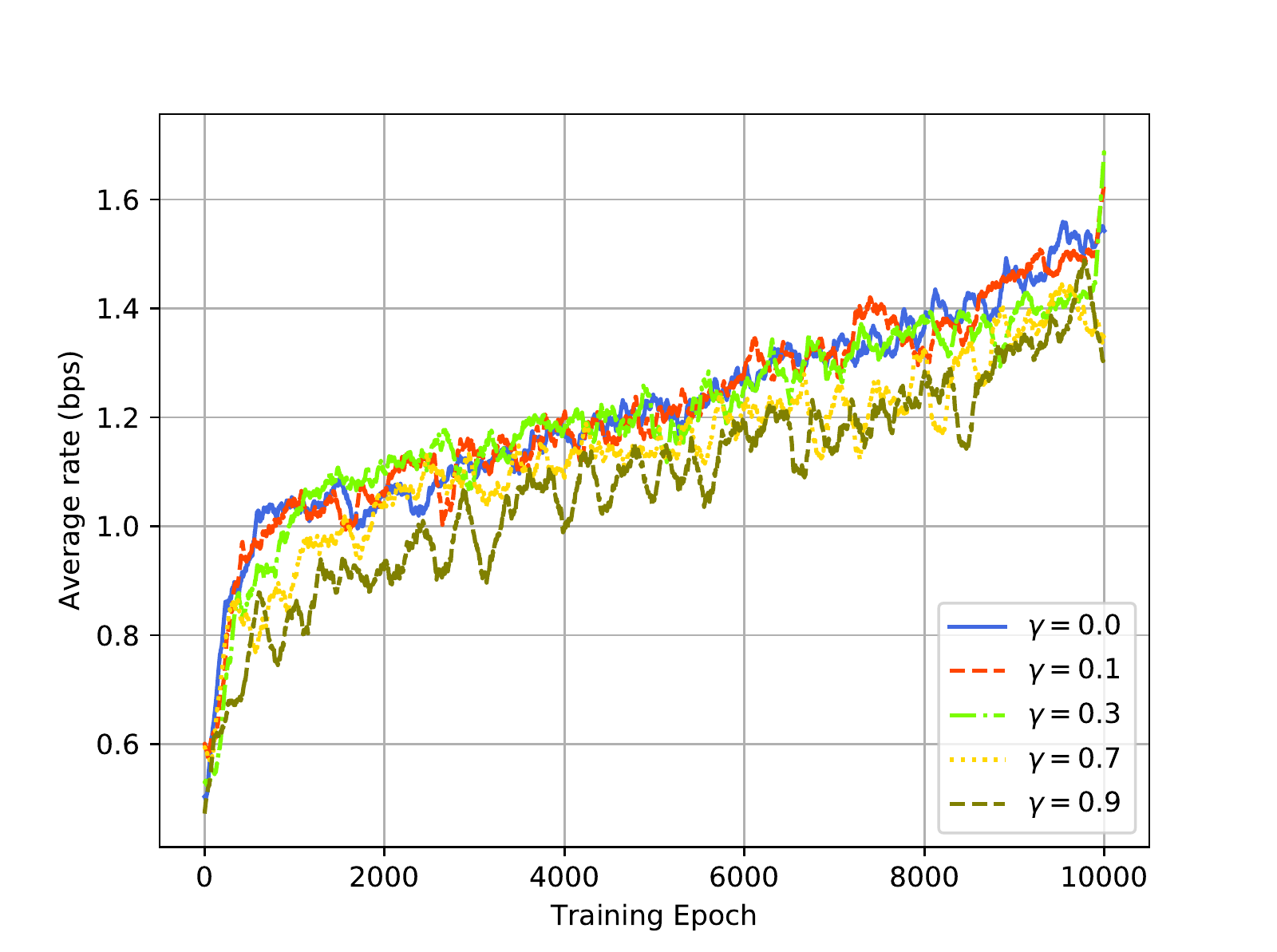}
	\caption{With different $ \gamma $ values, the recorded average rate during training period (Curves smoothed by averaged window).}
	\label{fig:gamma_train}
\end{figure}

\begin{figure}
	\centering
	\includegraphics[width=3.6in]{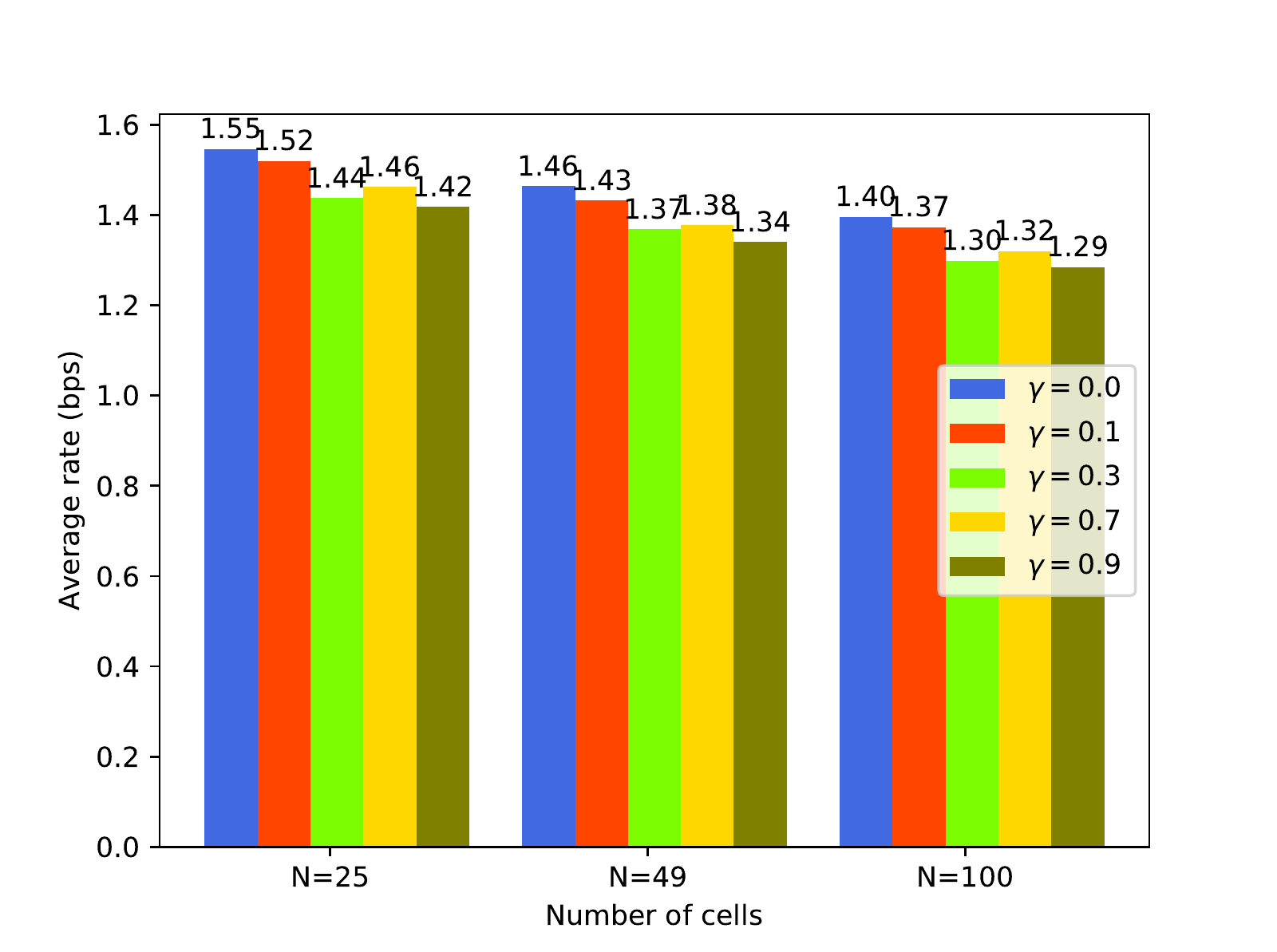}
	\caption{The average rate $ \bar{C} $ versus cellular network scalability for trained DQNs with different $ \gamma $ values.}
	\label{fig:gamma_test}
\end{figure}

In this subsection, the performance of different discount factor $ \gamma $ is studied. We set $ \gamma \in \{0.0,0.1,0.3,0.7,0.9\} $, and the average rate $ \bar{C} $ over the training period is shown in Fig.~\ref{fig:gamma_train}. At the same time slot, obviously the values of $ \bar{C} $ with higher $ \gamma \in \{0.7,0.9\} $ are lower than the rest with lower $ \gamma $ values. The trained DQNs are then tested in three cellular networks with different cell numbers. As shown in Fig.~\ref{fig:gamma_test} shows that DQN with $ \gamma = 0.0 $ achieves the highest $ \bar{C} $ score, while the lowest value is obtained by the one with highest $ \gamma $ value. The simulation result shows that the non-zero $ \gamma $ has a negative influence on the performance of DQN, which is consistent with the analysis in \ref{sec:dis}. Therefore, a zero or low discount factor value is recommended.

\subsection{Algorithm Comparison}\label{vs}

\begin{figure}
	\centering
	\includegraphics[width=3.6in]{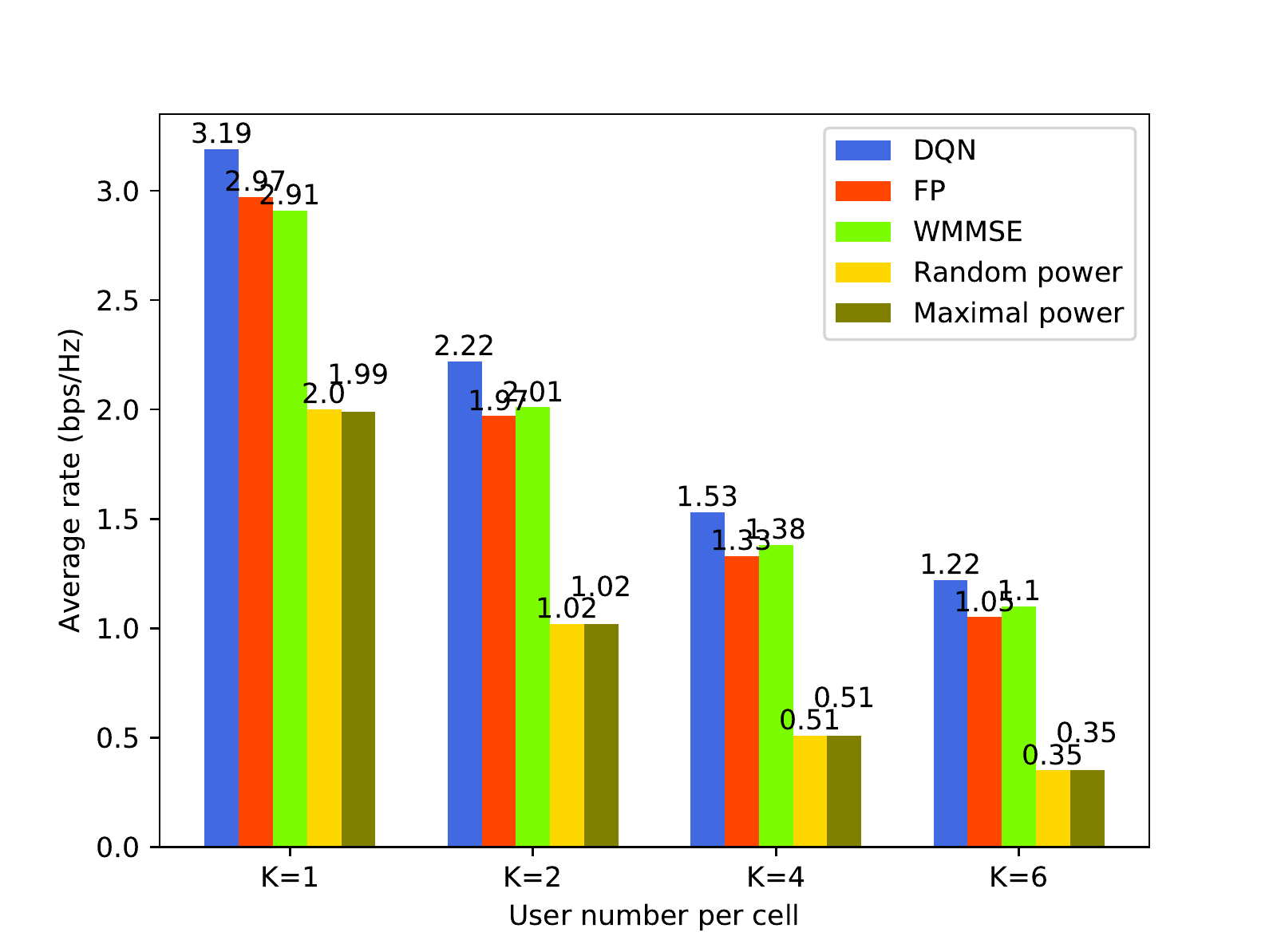}
	\caption{The average rate $ \bar{C} $ versus user number per cell. Five power allocation schemes are tested.}
	\label{fig:rate_user}
\end{figure}

The DQN trained with zero $ \gamma $ is used, and the four benchmark algorithms stated before are tested as comparisons. In real cellular network, the user density is changing over time, and the DQN must have good generalization ability against this issue. The user number per cell $ K $ is assumed to be in set $ \{1,2,4,6\} $. The averaged simulation results are obtained after $ 500 $ repeats. As shown in Fig.~\ref{fig:rate_user}, the DQN achieves the highest $ \bar{C} $ in all testing scenarios. Although it is trained with $ K=4 $, the DQN still outperforms the other algorithms in the other cases. We also note that the gap between random/maximum PA schemes and the rest optimization algorithms is increased when $ K $ becomes larger. This can be mainly attributed that the intra-cell interference gets stronger with increased user density, which indicating that the optimization of PA is more significant in the cellular networks with denser users.  

\begin{figure}
	\centering
	\includegraphics[width=3.6in]{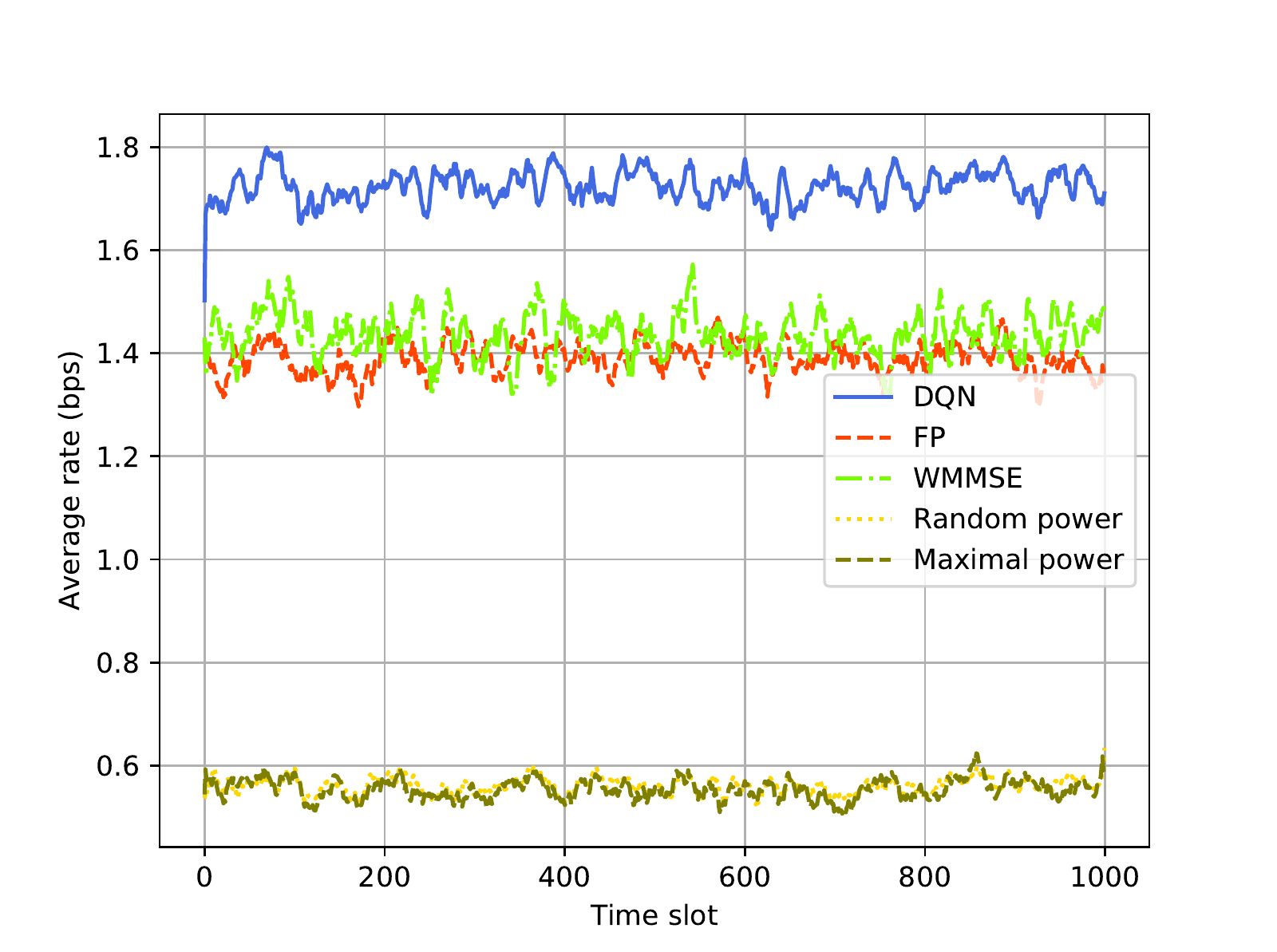}
	\caption{Comparisons of all five power allocation schemes over $ 1000 $ time slots (Curves smoothed by averaged window).}
	\label{fig:rate_time}
\end{figure}

We also give an example result of one testing episode here ($ K = 4 $). In comparison with the averaged sum-rate values in Fig.~\ref{fig:rate_user}, in Fig.~\ref{fig:rate_time} the performance of three PA algorithms (DQN, FP, WMMSE) is not stable, especially depending on the specific large-scale fading effects. Additionally, in some episodes the DQN can not be better than the other algorithms over the time (not shown in this paper), which means that there is still potential to improve the DQN performance. 

In terms of computation complexity, the time cost of DQN is in linear relationship with layer numbers, with the utilization of GPU. Meanwhile, both FP and WMMSE are iterative algorithms, and thus the time cost is not constant, depending on the stopping criterion condition, initialization and CSI.

\section{Conclusions}\label{sec:con}

The PA problem in the cellular network with IMAC has been investigated, and the data-driven model-free DQL has been applied to solve this issue. To be in consistent with the PA optimization target, the current sum-rate is used as reward function, including no future reward. This designed DQL algorithm is proposed, and the DQN simply works as an estimator to predict the current sum-rate under all power levels with a certain CSI. Simulation results show that the DQN trained with zero $ \gamma $ achieves the highest average sum-rate. Then in a series of different scenarios, the proposed DQN outperforms the benchmark algorithms, indicating that the designed DQN has good generalization abilities. In our two-step training framework, we have realized the off-line centralized learning with simulated communication networks, and the learned DQN is tested by distributed executions. In our future work, the on-line learning will be further studied to accommodate the real scenarios with specific user distributions and geographical environments.

\section{Acknowledgments}

This work was supported in part by the  National Natural Science Foundation of China (Grant No. 61801112, 61471117, 61601281), the Natural Science Foundation of Jiangsu Province (Grant No. BK20180357), the Open Program of State Key Laboratory of Millimeter Waves (Southeast University, Grant No. Z201804).

\input{revised.bbl}

\end{document}

%% file: revised.bbl

%% file: revised.bbl
\begin{thebibliography}{10}
\providecommand{\url}[1]{#1}
\csname url@samestyle\endcsname
\providecommand{\newblock}{\relax}
\providecommand{\bibinfo}[2]{#2}
\providecommand{\BIBentrySTDinterwordspacing}{\spaceskip=0pt\relax}
\providecommand{\BIBentryALTinterwordstretchfactor}{4}
\providecommand{\BIBentryALTinterwordspacing}{\spaceskip=\fontdimen2\font plus
\BIBentryALTinterwordstretchfactor\fontdimen3\font minus
  \fontdimen4\font\relax}
\providecommand{\BIBforeignlanguage}[2]{{%
\expandafter\ifx\csname l@#1\endcsname\relax
\typeout{** WARNING: IEEEtran.bst: No hyphenation pattern has been}%
\typeout{** loaded for the language `#1'. Using the pattern for}%
\typeout{** the default language instead.}%
\else
\language=\csname l@#1\endcsname
\fi
#2}}
\providecommand{\BIBdecl}{\relax}
\BIBdecl

\bibitem{Shen2018Fractional}
K.~Shen and W.~Yu, ``Fractional programming for communication systems—part i:
  Power control and beamforming,'' \emph{IEEE Transactions on Signal
  Processing}, vol.~66, no.~10, pp. 2616--2630, 2018.

\bibitem{Shi2011An}
Q.~Shi, M.~Razaviyayn, Z.~Q. Luo, and C.~He, ``An iteratively weighted mmse
  approach to distributed sum-utility maximization for a mimo interfering
  broadcast channel,'' in \emph{IEEE International Conference on Acoustics,
  Speech and Signal Processing}, 2011, pp. 4331--4340.

\bibitem{Chiang2008Power}
M.~Chiang, P.~Hande, T.~Lan, and C.~W. Tan, ``Power control in wireless
  cellular networks,'' \emph{Foundations and Trends in Networking}, vol.~2,
  no.~4, pp. 381--533, 2008.

\bibitem{5770666}
H.~Zhang, L.~Venturino, N.~Prasad, P.~Li, S.~Rangarajan, and X.~Wang,
  ``Weighted sum-rate maximization in multi-cell networks via coordinated
  scheduling and discrete power control,'' \emph{IEEE Journal on Selected Areas
  in Communications}, vol.~29, no.~6, pp. 1214--1224, June 2011.

\bibitem{Liye}
\BIBentryALTinterwordspacing
Z.~Qin, H.~Ye, G.~Y. Li, and B.~F. Juang, ``Deep learning in physical layer
  communications,'' \emph{CoRR}, vol. abs/1807.11713, 2018. [Online].
  Available: \url{http://arxiv.org/abs/1807.11713}
\BIBentrySTDinterwordspacing

\bibitem{8054694}
T.~O’Shea and J.~Hoydis, ``An introduction to deep learning for the physical
  layer,'' \emph{IEEE Transactions on Cognitive Communications and Networking},
  vol.~3, no.~4, pp. 563--575, Dec 2017.

\bibitem{Lecun2015Deep}
Y.~Lecun, Y.~Bengio, and G.~Hinton, ``Deep learning.'' \emph{Nature}, vol. 521,
  no. 7553, p. 436, 2015.

\bibitem{8444648}
H.~Sun, X.~Chen, Q.~Shi, M.~Hong, X.~Fu, and N.~D. Sidiropoulos, ``Learning to
  optimize: Training deep neural networks for interference management,''
  \emph{IEEE Transactions on Signal Processing}, vol.~66, no.~20, pp.
  5438--5453, Oct 2018.

\bibitem{8454504}
F.~Meng, P.~Chen, L.~Wu, and X.~Wang, ``Automatic modulation classification: A
  deep learning enabled approach,'' \emph{IEEE Transactions on Vehicular
  Technology}, pp. 1--1, 2018.

\bibitem{8052521}
H.~Ye, G.~Y. Li, and B.~Juang, ``Power of deep learning for channel estimation
  and signal detection in ofdm systems,'' \emph{IEEE Wireless Communications
  Letters}, vol.~7, no.~1, pp. 114--117, Feb 2018.

\bibitem{DBLP:journals/corr/abs-1803-06760}
\BIBentryALTinterwordspacing
R.~Amiri, H.~Mehrpouyan, L.~Fridman, R.~K. Mallik, A.~Nallanathan, and
  D.~Matolak, ``A machine learning approach for power allocation in hetnets
  considering qos,'' \emph{CoRR}, vol. abs/1803.06760, 2018. [Online].
  Available: \url{http://arxiv.org/abs/1803.06760}
\BIBentrySTDinterwordspacing

\bibitem{7997440}
E.~Ghadimi, F.~D. Calabrese, G.~Peters, and P.~Soldati, ``A reinforcement
  learning approach to power control and rate adaptation in cellular
  networks,'' in \emph{2017 IEEE International Conference on Communications
  (ICC)}, May 2017, pp. 1--7.

\bibitem{8466370}
F.~D. Calabrese, L.~Wang, E.~Ghadimi, G.~Peters, L.~Hanzo, and P.~Soldati,
  ``Learning radio resource management in rans: Framework, opportunities, and
  challenges,'' \emph{IEEE Communications Magazine}, vol.~56, no.~9, pp.
  138--145, Sep 2018.

\bibitem{8412128}
L.~Xiao, D.~Jiang, D.~Xu, H.~Zhu, Y.~Zhang, and V.~Poor, ``Two-dimensional
  anti-jamming mobile communication based on reinforcement learning,''
  \emph{IEEE Transactions on Vehicular Technology}, pp. 1--1, 2018.

\bibitem{DBLP:DRL}
\BIBentryALTinterwordspacing
Y.~S. Nasir and D.~Guo, ``Deep reinforcement learning for distributed dynamic
  power allocation in wireless networks,'' \emph{CoRR}, vol. abs/1808.00490,
  2018. [Online]. Available: \url{http://arxiv.org/abs/1808.00490}
\BIBentrySTDinterwordspacing

\bibitem{Mnih2015Human}
V.~Mnih, K.~Kavukcuoglu, D.~Silver, A.~A. Rusu, J.~Veness, M.~G. Bellemare,
  A.~Graves, M.~Riedmiller, A.~K. Fidjeland, and G.~Ostrovski, ``Human-level
  control through deep reinforcement learning.'' \emph{Nature}, vol. 518, no.
  7540, p. 529, 2015.

\bibitem{Introduction}
S.~Sutton and A.~G. Barto, \emph{Reinforcement Learning: An
  Introduction.}\hskip 1em plus 0.5em minus 0.4em\relax Cambridge, MA: MIT
  Press, 1998.

\bibitem{Kingma2014Adam}
\BIBentryALTinterwordspacing
D.~P. Kingma and J.~Ba, ``Adam: {A} method for stochastic optimization,''
  \emph{CoRR}, vol. abs/1412.6980, 2014. [Online]. Available:
  \url{http://arxiv.org/abs/1412.6980}
\BIBentrySTDinterwordspacing

\end{thebibliography}
